\documentclass{aa}  
\usepackage{showyourwork}
\newcommand\picb{$\beta$ Pictoris b}
\newcommand\picc{$\beta$ Pictoris c}
\newcommand\picd{$\beta$ Pictoris d}
\newcommand\pic{$\beta$ Pictoris}
\usepackage[para,online,flushleft]{threeparttable}
\usepackage{subcaption}

\newsavebox{\twosubbox}
\usepackage{lscape}
\usepackage{graphicx}
\usepackage{txfonts}

\usepackage{dblfloatfix}
\usepackage{color,soul}
\usepackage{hyperref}
\hypersetup{%
        colorlinks,
        breaklinks=true,
        plainpages=false,%
        citecolor=[rgb]{0,0.20,0.45},
        linkcolor=[rgb]{0,0.20,0.45},
        urlcolor=[rgb]{0,0.20,0.45},
        bookmarksopen=true,%
        bookmarksnumbered=false,%
        bookmarksdepth=5%
}
\usepackage{placeins}
\usepackage[nolist,nohyperlinks]{acronym}
\newacro{bring}[bRing]{\picb{} Rings}
\newacro{sphere}[SPHERE]{Spectro-Polarimetric High-contrast Exoplanet REsearch}

\makeatletter
\renewcommand*\aa@pageof{, page \thepage{} of \pageref*{LastPage}}
\makeatother

\begin{document} 
\authorrunning{Bakker et al.}
\titlerunning{\picc{} Hill sphere transits}
  \title{The Hill sphere transits of \picc{} \\ and the search for another 1981-like event}

\author{V.E.C. Bakker\inst{1}
\and
M.A. Kenworthy\inst{1}
\and
R. Stuik\inst{1}
\and
K. Zwintz\inst{2}
\and
T. Guillot\inst{3}
\and
S. Vach\inst{4,5}
\and
J. Wang\inst{6,7}
\and
J. Kammerer\inst{4}
\and
W. Balmer\inst{8,9}
\and
S. Lacour\inst{10,4}
\and
S. Zieba\inst{11}
\and
E.E. Mamajek\inst{12}
\and
G.-D. Marleau\inst{13,14,15}
\and
R.B. Kuhn\inst{16,17}
\and
P. Kalas\inst{18,19}
\and
Y. De Pra\inst{20}
\and
E.J.W. de Mooij\inst{21}
\and
N. Crouzet\inst{22,1}
\and
M. Buttu\inst{23}
\and
J. I. Bailey, III\inst{24}
}

\institute{Leiden Observatory, Leiden University, Postbus 9513, 2300 RA Leiden, The Netherlands
 \and
Universit{\"a}t Innsbruck, Institute for Astro- and Particle Physics, Technikerstra{\ss}e 25, 6020 Innsbruck, Austria
 \and
Observatoire de la C\^ote d'Azur, Laboratoire Lagrange CNRS UMR 7293, Universit\'e C\^ote d'Azur, 06304 Nice, France
 \and
European Southern Observatory, Karl-Schwarzschild-Stra{\ss}e 2, 85748 Garching, Germany
 \and
University of Southern Queensland, West Street, 4350 Toowoomba, Australia
 \and
Center for Interdisciplinary Exploration and Research in Astrophysics (CIERA), Northwestern University, Evanston, IL 60208, USA
 \and
Department of Physics and Astronomy, Northwestern University, Evanston, IL, USA
 \and
Department of Physics \& Astronomy, Johns Hopkins University, 3400 N. Charles Street, Baltimore, MD 21218, USA
 \and
Space Telescope Science Institute, 3700 San Martin Drive, Baltimore, MD 21218, USA
 \and
LESIA, Observatoire de Paris, PSL, CNRS, Sorbonne Universit\'{e}, Universit\'{e} de Paris, 5 place Janssen, 92195 Meudon, France
 \and
Center for Astrophysics | Harvard \& Smithsonian, 60 Garden Street, Cambridge, MA 02138, USA
 \and
Jet Propulsion Laboratory, California Institute of Technology, M/S321-162, 4800 Oak Grove Drive, Pasadena, CA 91109,USA
 \and
Max-Planck-Institut f\"{u}r Astronomie, K\"{o}nigstuhl 17, 69117 Heidelberg, Germany
 \and
Division of Space Research \& Planetary Sciences, Physics Institute, University of Bern, Sidlerstr. 5, 3012 Bern, Switzerland
 \and
Fakult\"{a}t f\"{u}r Physik, Universit\"{a}t Duisburg-Essen, Lotharstra{\ss}e 1, 47057 Duisburg, Germany
 \and
South African Astronomical Observatory, PO Box 9, Observatory, Cape Town, 7935, South Africa
 \and
Southern African Large Telescope, PO Box 9, Observatory, Cape Town, 7935, South Africa
 \and
Astronomy Department, University of California, Berkeley, CA, 94720 USA.
 \and
Institute of Astrophysics, Foundation for Research and Technology Hellas, Heraklion, 70013, Greece.
 \and
TERIN-DEC-H2V ENEA Italian National Agency for New Technologies, Energy and Sustainable Economic Development
 \and
Astrophysics Research Centre, School of Mathematics and Physics, Queen's University Belfast, Belfast BT7 1NN, UK
 \and
Kapteyn Astronomical Institute, University of Groningen, P.O. Box 800, 9700 AV Groningen, The Netherlands
 \and
Cagliari Observatory, Via della Scienza 5, 09047 Selargius (CA), Italy
 \and
Caltech Optical Observatories, California Institute of Technology, 1200 E California Blvd, Pasadena, CA 91125
 }

\date{Received 10 March 2026; accepted 27 July 2026}

\abstract{\pic{} is a young and nearby planetary system hosting an edge-on debris disk and at least three gas-giant planets.
Their circumplanetary environments could host exomoons and rings, a detection of which would be highly informative for moon and planet formation theories.
A photometric fluctuation of about 4\% was seen towards \pic{} in 1981, indicating the transit of dust in the system, which could be associated with the Hill spheres of the two inner planets.}
{We search for the origin of the 1981 event by searching for an analogous event in multi-epoch photometry from 2017 to 2023, and look for signs of circumplanetary material in transits of the Hill sphere of \picc{}.}
{Observations from the BRITE satellite, and the bRing and ASTEP observatories are fitted to a model of the 1981 event to search for a similar event, and also search for a signal consistent with a circumplanetary disk transit using a simple flat disk model.}
{No compelling evidence for a 1981 event during a Hill sphere primary transit is found, although we do find a candidate event around 2019 July 26.
Due to the uncertainty in the time of closest projected separation of \picc{}, a search for a disk during the primary transit of 2018 could not be robustly determined, but bRing photometry for the second primary transit places an upper limit on the dust content of the Hill sphere of $\sim 10^{22}$ grams of material.}
{With no compelling detection of an event similar to that seen in 1981 during the transits of \picc{} in the photometric time series, we rule out the hypothesis that this event was related to the Hill spheres of \picc{} and \picb{}.
Future observations of the next Hill sphere transit in 2028 can be realised with both ground-based observatories and with PLATO, whose first long duration science pointing will include \pic{}.
} 

\maketitle
\section{Introduction}
A circumplanetary disk (CPD) forms around a planet \citep{Taylor_2024} within its Hill sphere, defined as the radius where matter is gravitationally bound to the planet. 
In addition to dust and gas within the CPD being accreted onto the planets, it can dissipate by radiation pressure or the dust can collide and fragment \citep{Oberg_2020}.
The collisions can lead to dust clumping and eventually the creation of exosatellites, which continue to accrete material, creating gaps and rings.
In this paper we focus on two Hill sphere transits of \picc{}.

\pic{} (HD 39060, HIP 27321) is a bright ($V$=3.85), nearby ($d$=19.6\,pc), young ($\sim$23 Myr-old) star with an exoplanetary system of two known planets and a complex circumstellar disk (see Table 1; ESA 1997)  \citep{GaiaCollab23,Mamajek14,Smith_1984,Lagrange_2010}.
After the discovery of its edge-on debris disk \citep{Smith_1984}, it is one of the first disks to be spatially resolved, dominating the research field of debris disks \citep{Artymowicz_1997}.
One of the first major discoveries is the inner warped disk at 5$^\circ$ from the main disk \citep{Golimowski_2006}.
The primordial dust from the interstellar molecular cloud has likely been dispersed, leaving a system composed of disintegrating bodies in an early-phase planetary system \citep[e.g.][]{Backman_1993,Artymowicz_1997,Lagrange_2000,Zuckerman_2001}.
Moreover, star-grazing comets have provided evidence of their collisions and evaporation, giving rise to a mix of gas and dust in the disk \citep[e.g.][and references therein]{Lagrange_1998,Beust_1990,Lecavelier_Des_Etangs_1995,Beust_1996,Vidal_Madjar_2017}. 

The structure and composition of the disk led to the belief that the system hosts exoplanets, and in 2009 the first exoplanet in the system was detected by means of high-contrast imaging  \citep{Lagrange_2009}. 
The exoplanet \picb{} was found to have a mass of 11 $M_{\text{Jup}}$ and a period of 21 years orbiting at 9.8 au  \citep[e.g.][and many more]{Lagrange_2010, Chauvin_2012,Wang_2016,Snellen_2018,Kenworthy_2021}. 
Several years later \picc{} was discovered through the radial velocity technique  \citep{Lagrange_2019,Lagrange_2020,Nowak_2020}, where \citet{Lagrange_2020} found the exoplanet to be of comparable size to \picb{} with a mass of 7.8 $\pm$ 0.4 $M_{\text{Jup}}$ orbiting at closer proximity to the star with a semi-major axis of 2.7 $\pm$ 0.02 au. 
Astrometry of \picb{} was used to derive the astrometry of \picc{}, giving an approximate orbital period of 3.3 years \citep{Lacour_2021}. 
Most recently, a third gas giant planet, \picd{} has been simultaneously detected by two groups \citep{2026arXiv260623801S,2026arXiv260623789G} on a wide $\sim 25$ au orbit with a mass in the 2-4 ${\rm M_{Jup}}$ range.

\begin{figure}[t!]
 \centering
    \script{plot_m1981model.py}
    \includegraphics[width=0.48\textwidth]{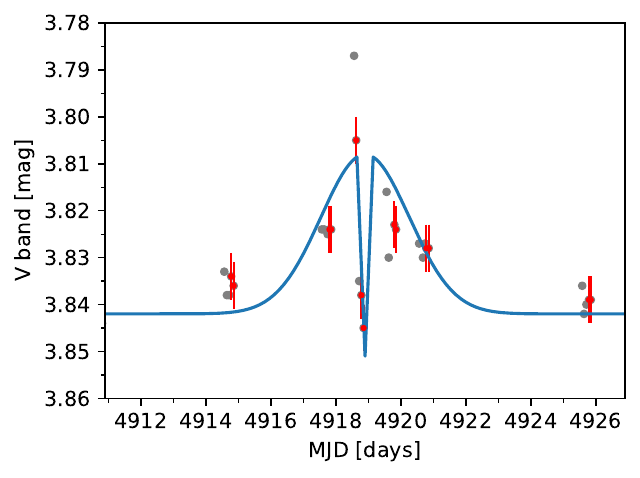}
    \caption{The data of the 1981 event with error bars (red).
    The grey dots did not pass the quality check \cite[][]{Lamers_1997,Lecavelier_Des_Etangs_1995}.
    The model (blue) consists of a broad, optically thin ring with a narrow, optically thick ring at its centre.
    The magnitude of the event is 0.035 \citep{Lamers_1997}.
    Credit: \citet{Kenworthy_2021}.}
    \label{fig:1981}
    
\end{figure}

On the 10th of November 1981 Universal Time (UT) the Geneva Observatory measured a significant increase in photometric flux of the \pic{} system \citep{ Lecavelier_Des_Etangs_1995}.
This mysterious `1981' event, shown in Fig.~\ref{fig:1981}, lasted 10 days with a dip halfway and detection in multiple colour bands with a confidence level of 99$\%$.
The origin of the event was strongly debated; one explanation could be caused by a giant comet \citep{Lamers_1997}, generating a cloud with highly forward-peaked scattering moving in front of the star \citep{Lecavelier_Des_Etangs_1995, Lecavelier_Des_Etangs_1997}.
Another explanation is a transiting planet with the following constraints, assuming a planet in circular motion \citep{Lecavelier_Des_Etangs_1997}: (i) the period is less than 19 years; (ii) the radius of the transiting object must be 2.3 to 4.0 times the radius of Jupiter.
The notion that the 1981 event was caused by a planet has become less favoured with the discovery of the planets in non-transiting orbits, where observations have shown that the large occultation depth is more likely to be produced by dust in a CPD \citep{Lecavelier_des_Etangs_2008a, Lecavelier_des_Etangs_2008b}. 
This last hypothesis of the 1981 event being caused by Rayleigh scattering from dust around an occulting planet can be tested. 
\picb{} and \picc{} both have a high enough orbital inclination that although they do not transit the stellar disk \citep{Wang_2016} their Hill spheres (possibly hosting a CPD) do transit the star to a significant depth.
Photometric data of the Hill sphere transit of \picb{} was fitted to the model of the 1981 event \citet[see Fig.~\ref{fig:1981}; ][]{Kenworthy_2021}, but no compelling results were found \citep{Lecavelier_Des_Etangs_2009,Kenworthy_2021}. 
With the discovery of \picc{}, the 1981 model from \citet{Kenworthy_2021} can be applied to \picc{} to see if a similar event occurred in photometric data.
We note that while the orbital fit of \picd{} \citep{2026arXiv260623801S} shows that its Hill sphere transits the star, no primary or secondary transit has occurred in the time-frames we are considering in this paper.

The Hill sphere transit gives additional insight into the CPD, as it lies within the Hill sphere and can host rings and moons, as seen with the solar system gas giants Saturn and Jupiter. 
However, outside of our solar system, only candidates for rings and exosatellites have been detected \citep{Osborn_2019,Kenworthy_Mamajek_2015,Mamajek_2012,Rieder_2016}. 
Simulations show detection from their light curve is possible \citep{Alshehhi_2020,Akinsanmi_2017} and our most confident candidate with a false positive probability of 1\% is the exomoon Kepler-1708 b-i around a Jupiter-sized exoplanet \citep{Kipping_2022}. 
Other candidates are hypothesized to host rings, such as PDS~110b, as observations have shown two eclipses transiting within a short timespan \citep{Osborn_2019}. 
Similarly, the J1407b ring system underwent multiple complex eclipses, possibly due to a substellar companion hosting rings \citep{Kenworthy_Mamajek_2015,Mamajek_2012,Rieder_2016}. 
In both cases, confirmation of exomoons and rings have not been successful. 
Improving our current instruments would allow a higher volume of direct imaging, which, for example, has allowed the detection and confirmation of the CPD around PSD 70c \citep{Benisty_2021}. 

The \pic{} system is young, giving us reason to believe its exoplanets are still hosting a CPD. 
\picb{} has been probed during its transit in 2017 and 2018 \citep{Kenworthy_2021} to fit for possible tilts and inclinations of the CPD extended to multiple radii and its mass limits. 
In this study, we search for transits of the hypothetical CPD for \picc{} using photometry from 2017 to 2023 following the methodology of \citet{Kenworthy_2021}.
We update the orbital parameters of the planets using GRAVITY astrometry in Sect.~\ref{sec:new_grav}.
After presenting the observations in Sect.~\ref{sec:Observations}, we carry out our analysis in Sect.~\ref{sec:Photo_Results} and discuss the results in Sect.\ref{sect:conc}.

\section{Updated orbital constraints}\label{sec:new_grav}

\subsection{New GRAVITY+ Observations}

We made use of three new GRAVITY observations and one new GRAVITY+ observation of the $\beta$~Pictoris system.
$\beta$~Pictoris b was observed on 2022 January 25 and 2022 August 18 as part of the ExoGravity program \citep[1104.C-0651;][]{Lacour_2020} using the GRAVITY instrument \citep{GRAVITY_Collaboration_2017} located on the Very Large Telescope Interferometer (VLTI) using the four 8-m Unit Telescopes (UT).
This program also observed $\beta$~Pictoris c on 2022 August 18.
Additionally, $\beta$~Pictoris c was observed as a part of a technical run on 2025 September 5 (TTR-115.0061) with the upgraded VLTI/GRAVITY+ \citep{GRAVITY+_Collaboration_2025} on the four UTs.
All observations were conducted using dual-field on-axis mode, with the fringe tracker observing the star, and the science fibre alternating between the host star and the predicted location as informed by previous astrometric observations \citep{Wang_2021}.\footnote{\texttt{whereistheplanet} is publicly available at \href{https://www.whereistheplanet.com}{https://www.whereistheplanet.com}} 

These data were reduced using the ESO GRAVITY pipeline v1.9.4 \citep{Lapeyrere_2014}.\footnote{The GRAVITY pipeline is available at \href{https://www.eso.org/sci/software/pipelines/gravity/gravity-pipe-recipes.html}{https://www.eso.org/sci/software/pipelines/gravity/gravity-pipe-recipes.html}} Following the procedure outlined in \citet{kammerer_2025}, we extracted the relative astrometry from each observation.
The extracted astrometry is presented in Table~\ref{tab:new_astrometry}.

\begin{table}
\caption{New relative astrometry of $\beta$~Pictoris~b and c. }
\label{tab:new_astrometry}
\centering
\begin{tabular}{l c c c}
\hline 
~~~ MJD & $\Delta$RA (mas) & $\Delta$Dec (mas) & $\rho$\\
\hline \hline \\[-0.3cm]
$\mathrm{\beta~Pictoris~b}$ & & & \\
~~~ 59604.160 & 256.753$\pm$0.056 & 422.165$\pm$0.141 & -0.499\\
~~~ 59809.384 & 273.776$\pm$0.067 & 447.813$\pm$0.060 & -0.770\\
\hline\\[-0.3cm]
$\mathrm{\beta~Pictoris~c}$ & & & \\
~~~ 59809.391 & 49.773$\pm$0.401 & 81.693$\pm$0.330 & -0.717\\
~~~ 60923.380 & 57.635$\pm$0.094 & 97.885$\pm$0.080 & -0.227\\
\hline
\end{tabular}
\tablefoot{
The observed astrometric offsets in right ascension ($\Delta$RA) and declination ($\Delta$Dec) are given in milli-arcseconds (mas). The correlation coefficient between right ascension and declination is given by $\rho$.}
\end{table}

\subsection{Orbit fit}

We combined the new relative astrometry observations presented above with the literature astrometry \citep{Lacour_2021, Nowak_2020}, imaging \citep{Lagrange_2020, Nielsen_2020}, radial velocity observations \citep{Lagrange_2020}, and absolute stellar astrometry \citep{Brandt_2021} to provide updated orbital parameters for both $\beta$~Pictoris~b and c using the \texttt{orbitize!} code \citep{orbitize}.
We fit a two-planet model to the data, fitting the system parallax, $\pi$, the stellar mass, $M_\star$, the planet masses, $M_b$ and $M_c$, and the orbital parameters of each planet: semi-major axis, $a$, orbital eccentricity, $e$, orbital inclination, $i$, argument of periastron, $\omega$, position of the ascending node, $\Omega$, and epoch of periastron in fractional orbital periods with respect to a reference epoch.
We used MJD 59000 as the reference epoch to allow for direct comparison against \citet{Lacour_2020}.
We imposed Gaussian priors on $M_\star$ informed by the literature measurements of $M_\star$ (see Table~\ref{tab:orbitize_param}), as well as the system parallax, using the same values as \citet{Lacour_2021}. 

To sample the posterior of estimated orbital parameters, we used the parallel-tempered affine-invariant sampler \texttt{ptemcee} \citep{Foreman-Mackey_2013, Vousden_2016} with 20 temperatures and 1000 walkers per temperature.
We discarded the first 4000 steps of every walker as a `burn-in' and only used the samples from the lowest temperature walkers for constructing our posterior.

\subsection{Geometry of the Hill sphere transit}\label{sec:Hill_sphere_transit}
\begin{figure}[t]
    \script{01_orbitalbundles_Beta_Pic_c.py}
    \centering
    \includegraphics[width=0.48\textwidth]{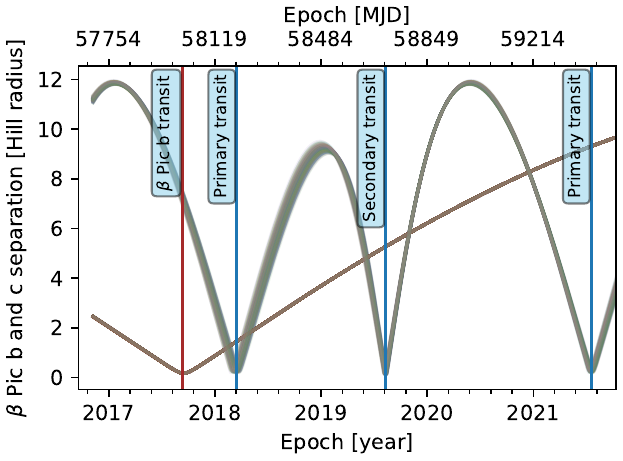}
    \caption{The separation of \picb{} and \picc{} from the star in terms of their Hill sphere.
    \picb{} has a longer orbital period ($\sim$20 yr), with a Hill sphere primary transit at MJD $\sim$58000 and a Hill sphere radius of 1.1 au. \picc{} has a shorter orbital period ($\sim$3.3 yr), with two primary transits at MJD $58195^{+11}_{-20}$ and $59415^{+8}_{-8}$ and a secondary transit at $58707^{+3}_{-7}$.
    The Hill sphere radius is approximately 0.3 au. }
    \label{fig:Orbitize_plot}
\end{figure}

The orbital bundles are processed in the Python \verb|Orbitize| model \texttt{orbitize!} using 200 orbits as shown in Fig.~\ref{fig:Orbitize_plot}.
The separation of \picb{} and \picc{} from the host star is shown in units of Hill radii \citep{Hamilton_1992}, for each respective planet using
\begin{equation}
    r_{\text{Hill}} \approx a \left(\frac{M_p}{M_{\star}+M_p} \right)^{\frac{1}{3}}  (1-e),
\end{equation}
where $a$ describes the semi-major axis between the star and the planet, $M$ are the masses and $e$ is the eccentricity. 
The parameters of the star and \picc{} are listed in Table \ref{tab:orbitize_param} and are used to obtain a Hill radius of 0.30 au. 

\begin{table}[t]
\caption{Adopted observational values for the \pic{} System and \picc{} extracted from references listed.}
\centering
\begin{tabular}{cccc} 
\hline
Parameter\,&Value\,&Units\,&Reference\\
\hline
\hline
$M_*$       & 1.797 $\pm$ 0.035     & $M_\odot$ & 1\\
$R_*$       & 1.497 $\pm$ 0.025     & $R_\odot$ & 1\\
$T_*$       & 8090 $\pm$ 59         & K         & 1\\
$L_*$       & 8.47 $\pm$ 0.23     & $L_{\odot}$ & 1\\
\hline
$M_c$       & 8.5 $\pm$ 0.5    & $M_J$     & 4\\
$R_c$       & 1.2 $\pm$ 0.1     & $R_J$     & 3\\
$T_c$       & 1250 $\pm$ 50      & K         & 3\\
$P_c$       & 1221 $\pm$ 15      & days         & 2\\
$a$          & 2.68 $\pm$ 0.02 & au       & 2 \\
$e$         &   0.208 $\pm$ 0.0074   &       & 4 \\
\hline
\end{tabular}

\tablebib{ (1) ~\citet{Zwintz_2019};
(2) \citet{Lacour_2021};
(3) \citet{Nowak_2020}; (4) \citet{Vandal_2020}.
}
\label{tab:orbitize_param}

\end{table}

The midpoints of the transits occur at the time of closest approach, where the Hill sphere passes in front of the star during the primary transit and behind the star during the secondary transit. 
Fig.~\ref{fig:Orbitize_plot} shows one full orbit of \picc{}, with two primary transits and one secondary transit. 
The most likely midpoint of the transits is derived from the mode of 200 orbits, with limits set by the earliest and latest timestamps of minimum separation. 
The computed primary transits are MJD $\text{58195} \substack{+\text{11} \\ -\text{20}}$ and $\text{59415} \substack{+\text{8} \\ -\text{8}}$ or 2018 March 18 and 2021 July 20, respectively. 
The secondary transit at MJD $\text{58707}\substack{+\text{3} \\ -\text{7}}$ (2019 August 12) has the lowest orbital uncertainties, resulting in the smallest dispersion.
The dispersion increases at timestamps further away from the secondary transit.

\section{Observations}\label{sec:Observations}

\begin{figure*}[t]
    \script{05_Photometric_data_transits_plots.py}
    \centering
    \includegraphics[width=0.9\textwidth]{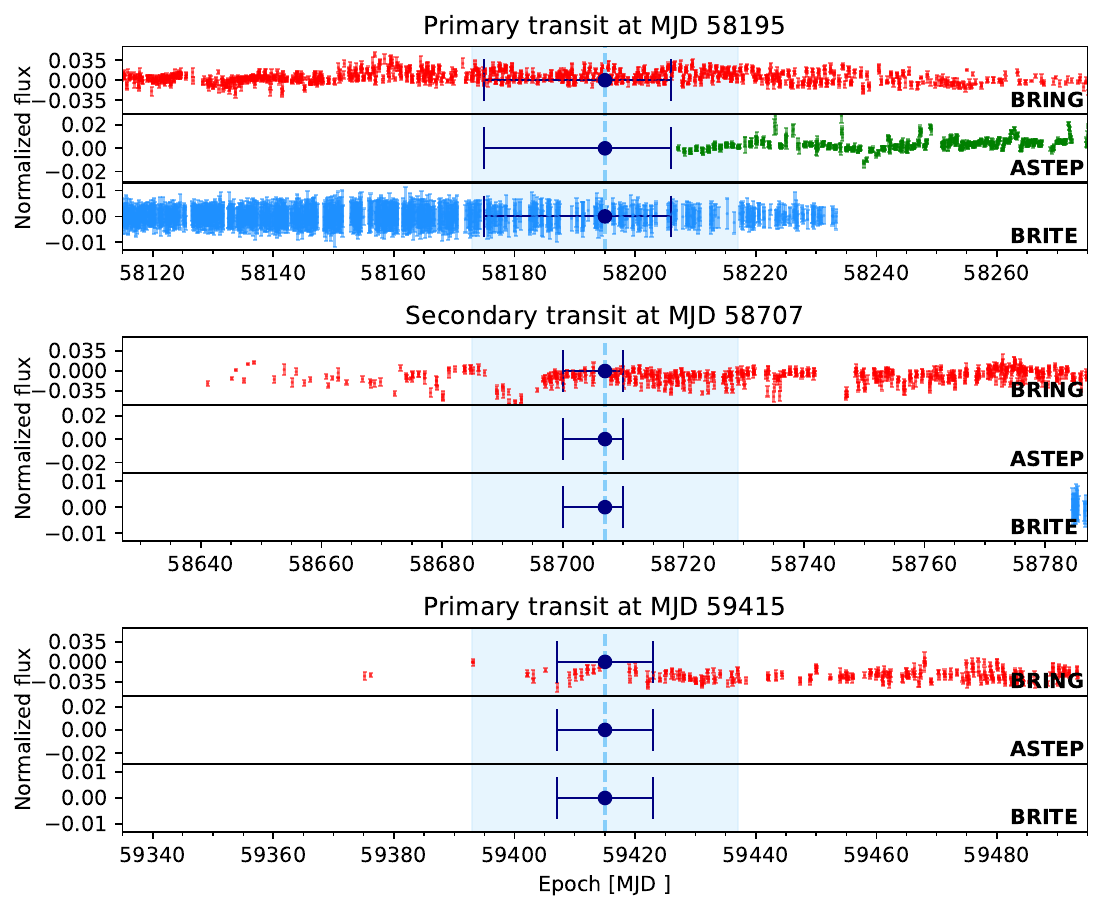}
    \caption{The binned photometric data of all Hill sphere transits of bRing, BRITE and ASTEP.
    The latter two only cover the primary transit at MJD 58195.
    The midpoints of the transits and their uncertainty are marked with an error bar and the blue boxes mark the duration of the transit ($\sim$1.5 months).
    No significant decrease in flux is seen over the three transits. }
    \label{fig:photometric_transits}
\end{figure*} 

The Hill sphere campaign of \picb{} \citep{Kenworthy_2021} observed the \pic{} system at high photometric cadence during the Hill sphere transit of \picb{}.
Close thereafter, \picc{} was discovered and the campaign coincidentally covered the transit of \picc{} as well \citep{Stuik_2017}. 
One of the instruments used was the \pic{} b Ring project (bRing) with almost full coverage of the system \citep{Snellen_2012,Snellen_2013,Stuik_2014, Stuik_2016,Talens_2017,Talens_2018}.
This instrument's coverage was further supported by Mascara and DREAM from 2017 February 03 to 2023 January 17.
The BRITE constellation \citep[BRIght Target Explorer; ][]{Weiss_2014} observed \pic{} from low Earth orbit, allowing higher precision than ground-based observations and negligible noise from atmospheric turbulence.  
The data quality of the blue filter was deemed insufficient for \pic{} \citep{Zwintz_2019} and only the red filter (550 -700 nm) was used. 
The data reduction pipeline is detailed in \citet{Popowic_2017} and a correction for \pic{} is given in \citet{Zwintz_2019}. 
The data from BRITE-Heweliusz (BHr) and BRITE-Toronto (BTr) cover 2015 March 16 to 2021 March 28 with irregular observations. 
The last instrument is ASTEP \citep{Crouzet_2010}, operational during the Antarctic winters and modified with a filter to observe the \pic{} system \citep{Kenworthy_2021}. 
ASTEP obtained data from 2017 March 30 to 2018 August 17.

The midpoints of the transits and their coverage are shown in Fig.~\ref{fig:photometric_transits}. 
The data has been normalised and binned to 0.05 days.
The error bars mark the uncertainty of the midpoint of the transits and the boxes mark the duration of the transit ($\sim$1.5 months). 
bRing covers all three transits, BRITE covers the first primary transit at MJD 58195 and ASTEP covers the tail end of this Hill sphere transit.
To probe the environment within the Hill sphere, a dimming of photometric flux due to dust within the Hill sphere is required and not scattered light, therefore, despite the small dip at MJD 58690 and the smallest uncertainty of the midpoint of the transit, the secondary transit does not give insight into the CPD.
The transit at MJD 58195 has the best coverage, however also the largest uncertainty in its midpoint.
The second primary transit at MJD 59415 has the least amount of photometric coverage and both transits show no significant flux deviations. 
\section{Photometric results and analysis}\label{sec:Photo_Results}
\subsection{The 1981 event}\label{sec:method_1981}

\begin{figure*}[t] 
    \script{06_1981_fit.py}
    \centering
        \includegraphics[trim={0 27.9cm 0 0},clip,width=0.9\textwidth] {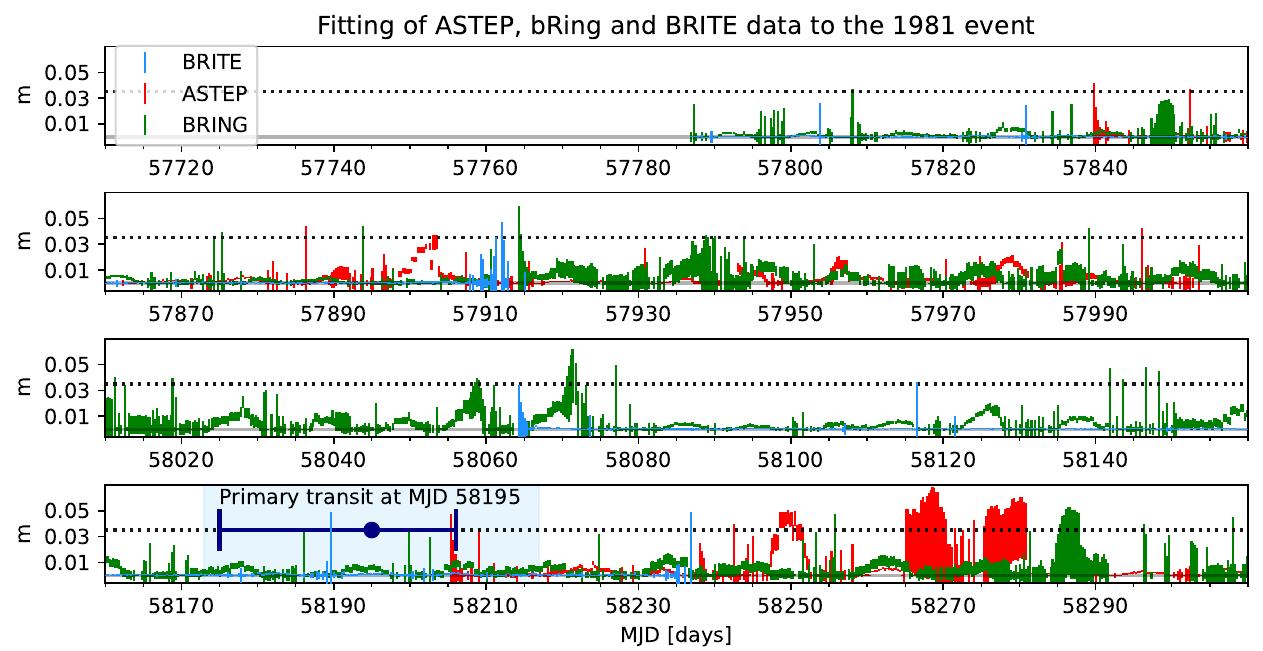} 
         \caption{The amplitude $m$ of the BRITE (blue), bRing (green) and ASTEP (red) data fitted to the 1981 event are shown, where 0.035 marks the magnitude of the 1981 event.
         The midpoints of the primary (MJD 58195 $\&$ 59415) and secondary (MJD 58707) Hill sphere transits and their uncertainty are marked with an error bar and the blue boxes marks the duration of the transit ($\sim$1.5 months). The figure is continued in Fig.~\ref{fig:cross_correlation2}. }
         
     \label{fig:cross_correlation}
\end{figure*}

The light curve of the 1981 event (Fig.~\ref{fig:1981}) is modelled as two rings, where an optically thick but narrow ring lies within a broader, optically thin one, as described in \citet{Lamers_1997}. 
The forward scattering from the optically thin ring passing in front of the star is characterised by a Gaussian feature with a FWHM of 3.2 days and a magnitude of 0.035 \citep{Lamers_1997}.
Furthermore, the segment of the ring transiting is approximated as a straight line, as the radius of the ring is much larger than the diameter of the star.
The 1981 event is fitted to the photometric data following \citet{Kenworthy_2021} and the model $F_{1981}(t)$ shown in Fig.~\ref{fig:1981} is fitted for amplitude $m$ and offset $b$ using, 
\begin{equation}
F(t) = mF_{1981}(t_{mid}) + b.
\end{equation} 
The function returns the best fit $m, b$ and $\sigma_m$, where a compelling detection would yield a magnitude of the same height as the 1981-event ($m$=0.035). 
Larger amplitudes than 0.05 are discarded and $F(t)$ masks out any data points more than 8 days away from $t_{mid}$, fitting only for the duration of the 1981 event.
All observational data from ASTEP, bRing and BRITE were fitted to the model, shown in Fig.~\ref{fig:cross_correlation} and Fig.~\ref{fig:cross_correlation2}.
bRing has the best coverage, while ASTEP's coverage is limited to the first 500 days, and BRITE's coverage is very sporadic.  
The transits of $\sim$1.5 months are marked with a blue box, while the uncertainties in the midpoints of the eclipse are marked by error bars.
The magnitude of the 1981 event is indicated by a horizontal line with $m$=0.035. 
A good fit to the model would yield a smooth, symmetric increase and decline in amplitude to 0.035 during 8 days over multiple instruments.
The latter is unfortunately difficult to achieve due to a lack of coverage from ASTEP and BRITE. 

No such ideal fit is seen, however multiple epochs exceed the 0.035 threshold.
The primary transit at MJD 58195 does not show a compelling detection, while the second primary transit at MJD 59415 does if taking the error of the midpoint into account. 
The errors among the amplitudes $m$ of the fit should be taken conservatively, where at MJD 59415 the detection threshold is just met.
The increase and decrease last around $\sim$9 days with a non-symmetric curve due to the outliers seen at MJD $\sim$59417. 
In addition, ASTEP and BRITE can not confirm this detection and it is not seen during the first primary transit. 
As a result, the MJD 59415 detection can not be confidently associated with the 1981 event.
The secondary transit at MJD 58707 shows a detection (MJD 58690) just outside the expected transit times, however the secondary transit requires a different fitting model incorporating backscattering, therefore the detection at MJD 58690 can not be attributed to the secondary transit. 
The MJD 58690 detection could be caused by other unknown events.
The remaining detections ($m$>0.02) are short-lived for about a day or two, likely not a 1981-like event.
There is no strong evidence that the 1981 event was caused by the Hill sphere transit of \picc{}.  

Fig.~\ref{fig:cross_correlation} (continued in Fig.~\ref{fig:cross_correlation2}) shows additional features outside the transits. 
A symmetric curve at MJD 58250 around the threshold of 0.035 is seen with ASTEP, however bRing and BRITE do not show this feature. 
Similarly at MJD 57900, ASTEP observed an increase in flux, while BRITE and bRing do not, possibly due to systematic errors or interruptions in telescope operation \citep{Kenworthy_2021}.
Furthermore, large errors at MJD 58270 are seen with ASTEP, therefore we consider these detections unreliable. 
At MJD 58070, MJD 58280, MJD 58750 and MJD 58930 bRing detects a signal, however the photometric errors are often large and are not consistent with the other instruments. 
BRITE has the most accurate data as a space-based telescope and shorter observation times due to its orbital period around the Earth and sees no significant detections.

Out of the Hill sphere transit times there are quasi-periodic variations with an amplitude of $m$ $\sim$0.01 over a span of 5-15 days, which occasionally overlap with other instruments (MJD 57960), however not always (MJD 57980).
Several mechanisms could create brightness pulsations in the \pic{} system, where many are on the order of minutes to hours.
The star has $\delta$ Scuti pulsation modes with a period of 26-30 minutes and a magnitude of $\leq$ 1.5 mmag \citep{Koen_2003}, neither of which can explain what we see.
Moreover, exocomets can also not cause such a feature, because they are short-lived, typically lasting less than two days, are asymmetric and of small amplitudes \citep[$<$2 mmag;][]{Heller_2024, Zieba_2019,  Pavlenko_2022, Lecavelier_2022}.  
It is notable that BRITE does not see such large magnitude events, suggesting ground-based systematics from atmospheric glow or contamination from scattered light from the Moon. 

We see a candidate for the 1981-like event at MJD 58690 (2019 July 26) with an unknown event origin and no compelling evidence during the primary transits.
Other transiting bodies could have created the 1981 event, however we are not aware of any other bodies large enough to do so in the system.  
Exocomets are too short-lived and not bright enough.  
Another hypothesis is forward scattering by a dust cloud, with dust clouds several times the diameter of the star, resulting in brightness variations at longer timescales.  
At the right tilt and inclination, a dust cloud transit could have shorter timescales, and possibly overlap with the epoch of the 1981 event.  
Furthermore, large comets can create dust clouds, which either pass the line of sight at the perihelion or are fragmented and collide with the star \citep{Lamers_1997}, creating so-called Shoemaker--Levi fragments: where a comet is captured into a satellite orbit around an exoplanet and eventually broken apart by tidal disruption events. 
The fragments collide with the exoplanet creating plumes and therefore an increase in flux.
In conclusion, the origin of the 1981 event remains unknown and we do not see the 1981 event in our photometric analysis during the primary Hill sphere transits.
The hypotheses about a transiting planet or its Hill sphere have become more unlikely, as no compelling evidence was found with \picb{} and \picc{} \citep{Lecavelier_Des_Etangs_1997, Kenworthy_2021}.
\subsection{The circumplanetary disk}\label{sec:method_CPD}

\begin{figure*}[!t]
    \script{07_Sensitivity_CPD_model.py}
    \centering
    \includegraphics[trim={0 0 0 0},clip,width=0.9\textwidth]{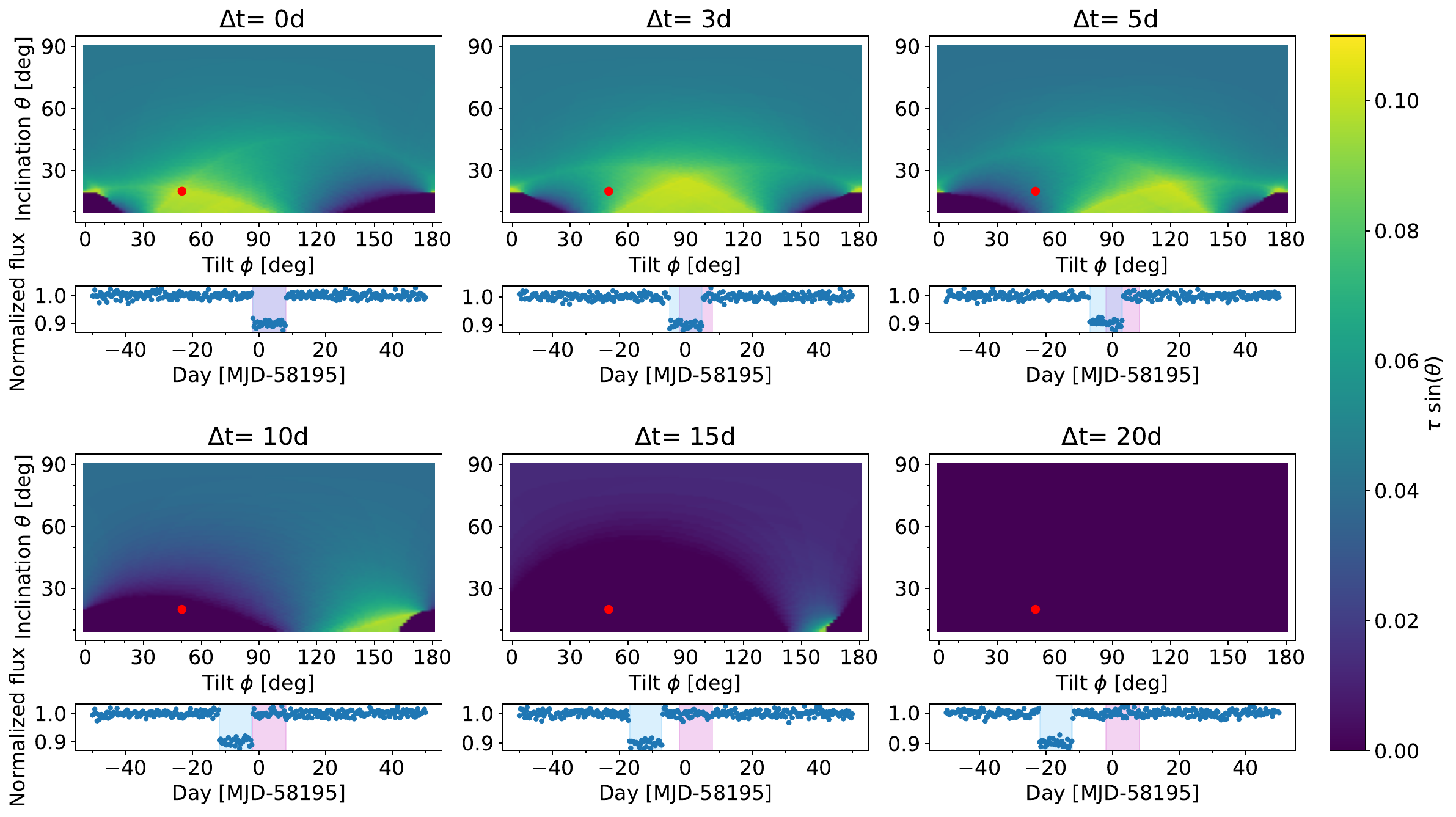}
    \caption{The sensitivity of the CPD model given a Hill sphere transit midpoint (MJD 58195) with an offset of $\Delta t$ relative to the real midpoint (MJD 58195 $\pm$ $\Delta t$). 
    A model of the Hill sphere transit with a disk inclined at 20\textdegree\ and tilted at 50\textdegree\ was used from \citet{Kenworthy_2021}. 
    The photometry of the transit is shown in the lower panel of each subplot, with an offset ranging from 0 to 20 days. 
    The maximum offset is based on the uncertainty of the \picc{} transit (MJD $58195^{+11}_{-20}$). 
    Each lower subplots has a blue box marking the transit times with an offset and a purple box for the transit times without offset.
    The model fits the photometry for a given midpoint (MJD 58195) and returns the optical depth for a grid of inclinations and tilts (upper subplots). 
    The highest optical depth corresponds with the most likely disk orientation, marked with the red dot for our model. 
    The accuracy of the model retrieving the correct disk orientation degrades rapidly with $\Delta t$.
}
    \label{fig:model_timestamps}
\end{figure*}  

The CPD model probes the environment within the Hill radius to determine possible disk radii, disk orientation, and the upper mass limit.
We follow the methods of \citet{Kenworthy_2021} and summarize them here.
To understand the assumptions made for the model, we want to determine whether circumplanetary material around \picc{} is rocky or icy. 
The equilibrium temperature of dust $T_g$ with a Bond albedo $A_V$ at a distance $D$ from a blackbody with radius $R$ and temperature $T$ is:

\begin{equation}
    T_g = \sqrt{\frac{R}{2D}}T (1-A_V)^\frac{1}{4}.
\end{equation}
For \picc{}, the thermal temperature from the star at the edge of the Hill sphere is 352 K for perfectly absorbing material, changing to 260 K for an albedo of 0.7 (comparable to Saturn's rings) using parameters from Table \ref{tab:orbitize_param} and the eccentricity of its orbit ($D = a[1-e]-r_{\text{Hill}}$).  
The thermal emission of the planet is expected to drop significantly at larger radii, therefore a temperature of 120 K is derived at 0.1 $r_{\text{Hill}}$.
The stellar contribution therefore dominates the radiation field, allowing us to assume that water ices would have sublimated rapidly and therefore silicates dominate the composition of any circumplanetary material, with a density of $\rho_g =\text{2.5 g/cm}^3$ \citep{Chen_Jura_2001,Li_1997}.
We say that the primordial disk material from \pic{} has likely been dispersed, resulting in a low abundance of circumplanetary gas.  
Catastrophic collisions within the inter-disk replenish the gas and dust in the environment, resulting in an abundance of small grains dominating the optical depth.
Radiation pressure from the star clears out the smallest particles from around the planet \citep{Kennedy_2011}, resulting in a minimum grain size of:
\begin{equation}
    D_{\text{min}} = 33 \left(\frac{r_{\text{\text{CPD}}}}{r_{\text{Hill}}}\right)^{1/2}  \mu \text{m},
    \label{eq:Mass}
\end{equation}
Where $r_{Hill}$ is the radius of the Hill sphere and $r_{CPD}$ the radius of the CPD.
The equation was derived from Eqn. 9 of \citet{Kennedy_2011} using parameters from Table \ref{tab:orbitize_param} and substituting the fractional radius $\eta = r_{\text{CPD}}/r_{\text{Hill}}$.
From here, a minimum grain size $D_{\text{min}}$ of 27 $\mu m$ is found, using the assumption that the dust is concentrated at the area mean weighted planetocentric distance of 0.7 $r_{\text{CPD}}$.
This is of similar order to the grain size of \picb{} and a couple of orders larger than the blow-out grain size \citep{Kenworthy_2021}, therefore the long-term presence and retention of dust in the CPD is plausible. 

The CPD model requires the grain size as one of the input parameters in addition to the photometric data and returns the optical depth given a grid of inclinations and tilts for the given disk radius, where the upper limit of the optical depth is used to compute the upper mass limit. 
The model approximates the disk as a thin homogeneous slab, with a height much smaller than the radius of the disk $r_{\text{CPD}}$. 
It assumes an optically thin slab to determine the optical depth,
\begin{equation} 
\tau = \frac{1-I/I_0}{\text{sin}(\theta)}.
\end{equation} 
Where $I_0$ is the attenuated star flux and $I$ is the observed star flux. 
The optical depth $\tau$, $r_{\text{CPD}},i,\phi,\text{t}_b$ and $t$ are used to compute the light curve $I(t)=f(r_{\text{CPD}},\tau,i,\phi,\text{t}_b$,t) by injecting a transit for $r \leq r_{\text{CPD}}$. 
A grid of trial inclinations $i$ and tilts $\phi$ are explored, as the obliquity of \picc{} is unknown and the obliquity of planets are unconstrained to their orbital plane \citep{Laskar_1993, Kenworthy_2021}.  
$t_b$ is the midpoint of the transit when the star passes behind the CPD at an impact parameter of the Hill sphere $b$ above the planet.
The impact parameter $b$ is expressed as a fraction of the $r_{\text{Hill}}$ and depends on the mass of the planet, as well as the orbital parameters.  
Since \picb{} and \picc{} have roughly the same masses and inclinations \citep{Kenworthy_2021}, the impact parameter is similar ($b$=0.1).  
However, it is worth emphasising that the impact parameter is not well constrained and could deviate by up to $\approx$10\% \citep[e.g.][]{Wang_2016,Lagrange_2019,Nielsen_2020}.  
Lastly, the truncation of the CPD disk $r_{\text{CPD}}$ can depend on the tilt of the CPD.
For a coplanar CPD, truncation happens at 0.4 $r_{\text{Hill}}$ when the tidal forces are too strong \citep{Martin_2011}.  
For non coplanar CPD, truncation can occur at different Hill radii.  
Since the orientation of the CPD is not known, the two most stable orbits are at 0.3 and 0.6 for prograde and retrograde motion respectively $r_{\text{Hill}}$ \citep{Lubow_2015, Miranda_2015}. 
The data for the primary transits (MJD 58195 \& MJD 59415) from each instrument is input into the model to determine $\tau$ within the inclination and tilt grid space and its $\chi^2$. 
Once the upper $\tau$ limit is found, the total mass of the dust can be computed using, 
\begin{equation}
    M_{\text{CPD}} = \frac{4 \tau \rho D_{\text{min}} }{3} \pi r_{\text{CPD}}^2,
    \label{eq:mass_limit}
\end{equation} 
Where the fraction represents the surface density $\sigma_{\text{CPD}} = \tau / \kappa$ with an opacity of $\kappa = 3/(4 \rho D_{\text{min}})$. 

\subsubsection{Sensitivity of the CPD model}

The midpoints of the Hill sphere transits have uncertainties up to 20 days at MJD 58195, therefore understanding the sensitivity of the CPD model to parameter uncertainty is crucial.
We modelled a disk with a radius of 0.6 $r_{\text{Hill}}$, an inclination of 20$^\circ$, a tilt of 50$^\circ$, an impact parameter of 0.2, an optical depth $\tau$ of 0.1 and a midpoint at MJD 58195.
The top left of Fig.~\ref{fig:model_timestamps} shows the results of minimising the optical depth using $\chi^2$ and shows two partial high optical depth paths intersecting at the best fit disk orientation, confirming the model can detect the tilt and inclination of the disk. 
We now add an offset to the midpoint of the transit without adjusting the given midpoint to the model (MJD 58195) and repeat the fit for offsets of [3,5,10,15,20] days.
Fig.~\ref{fig:model_timestamps} shows the optical depths in the upper panel of each subplot with a red dot marking the disk's parameters and the lower panel showing the model's photometric data. 
In blue the transit times with offset is highlighted and in purple the transit times at zero offset. 
The model is very sensitive to the given midpoint of the transit and three days is enough for the disk to be detected but with the incorrect parameters. 
Thus, the large uncertainties of the Hill sphere transits are the largest obstacle in fitting the model.

\begin{figure*}[!t]
\script{08_CPD_model_taumass_58195.py}
\centering
   \includegraphics[width=1\linewidth]{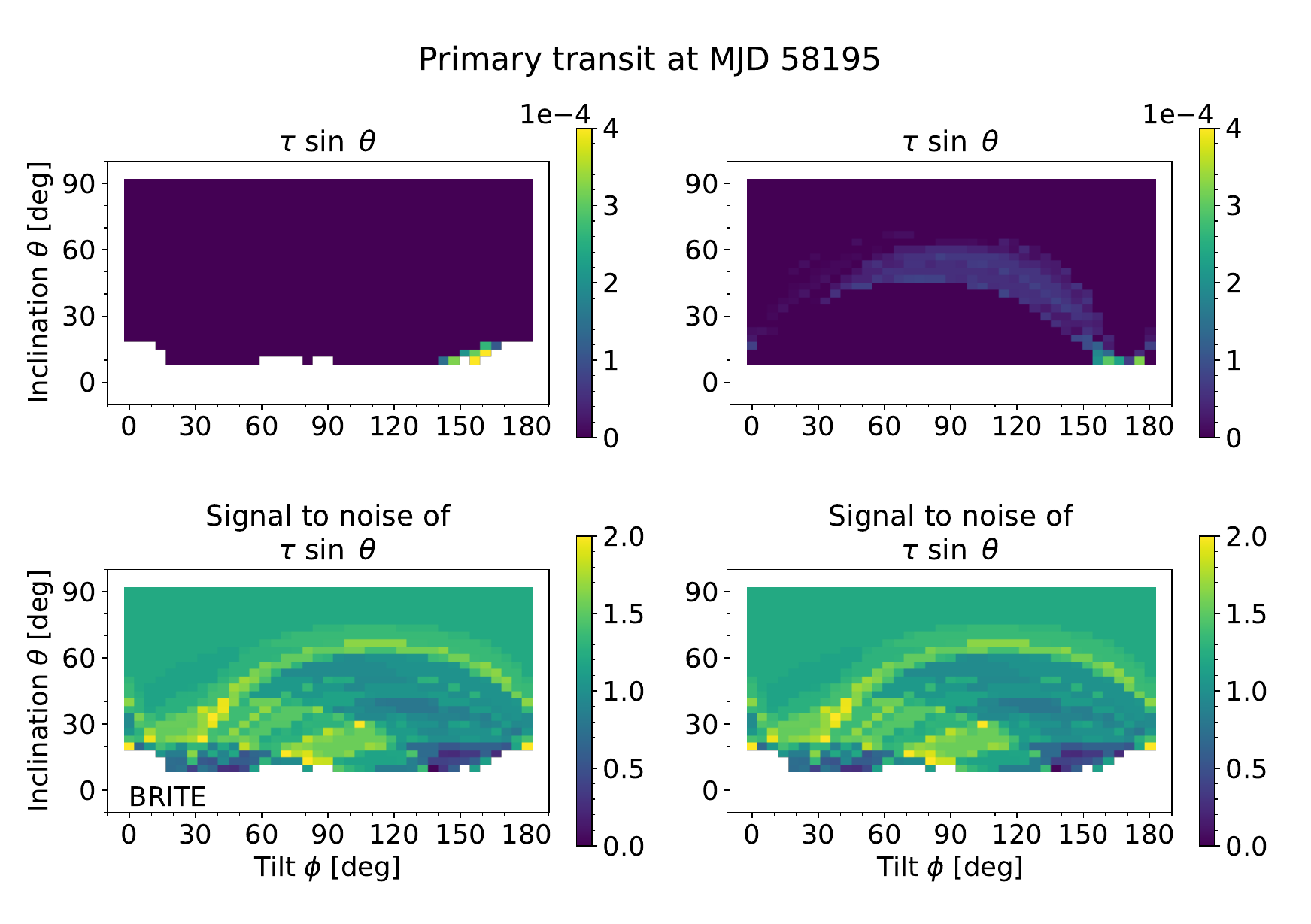}
   \label{fig:58195diskfit_3datasets_taumass_030}
\caption{The optical depth (upper subpanel) and SNR (lower subpanel) for a CPD fit at varying tilts and inclinations for the Hill sphere transit at MJD 58195 and a circumplanetary radius of 0.3 $r_{\text{Hill}}$ (upper panel) and 0.6 $r_{\text{Hill}}$ (lower panel). 
White mark the areas with no photometry to constrain the fit. 
Only BRITE is shown, as ASTEP has no coverage (Fig.~\ref{fig:photometric_transits}) and bRing contains unfeasible negative optical depth values. 
A confident CPD orientation corresponds to a single upper optical depth limit with a high SNR as seen in the upper left plots of Fig.~\ref{fig:model_timestamps}. 
The CPD fit at 0.3 $r_{\text{Hill}}$ has no upper optical depth and the fit at 0.6 $r_{\text{Hill}}$ shows a loosely constrained disk orientation with low SNR. }
\label{fig:58195diskfit_3datasets_taumass}
\end{figure*}
\subsubsection{Analysis of the CPD}

Fig.~\ref{fig:58195diskfit_3datasets_taumass} shows the optical depths at the two most stable CPD radii, namely 0.3 and 0.6 $r_{\text{Hill}}$, for the first primary transit at MJD 58195.
Only the CPD fit for BRITE is shown, because ASTEP has no coverage and bRing gives negative optical depth values.
bRing provides no constraints to the possible CPDs, because we assume the negative optical depths are due to systematic errors in its photometry, given that we do not see this effect in BRITE.
Among all the instruments and different CPD radii, no single upper $\tau$ limit can be found and therefore no single disk orientation.
At 0.3 $r_{\text{Hill}}$ no constraints on the CPD disk can be imposed.
More sensitive upper optical depth values are seen at 0.6 $r_{\text{Hill}}$, however they unfortunately correspond to low SNR values.
In addition, this pattern produced by the different optical depths within the grid of disk orientations (upper subplot at 0.6 $r_{\text{Hill}}$) is inconsistent with the way we expect the model to act with a midpoint offset according to Fig. \ref{fig:model_timestamps}. 
As a result, an offset of a few days is unlikely to yield a better fit.
There are too many uncertainties to constrain the tilt and inclination for a possible CPD at MJD 58195.

\begin{figure*}[t] 
\script{08_CPD_model_taumass_59415.py}
    \centering
    \includegraphics[width=0.9\textwidth]{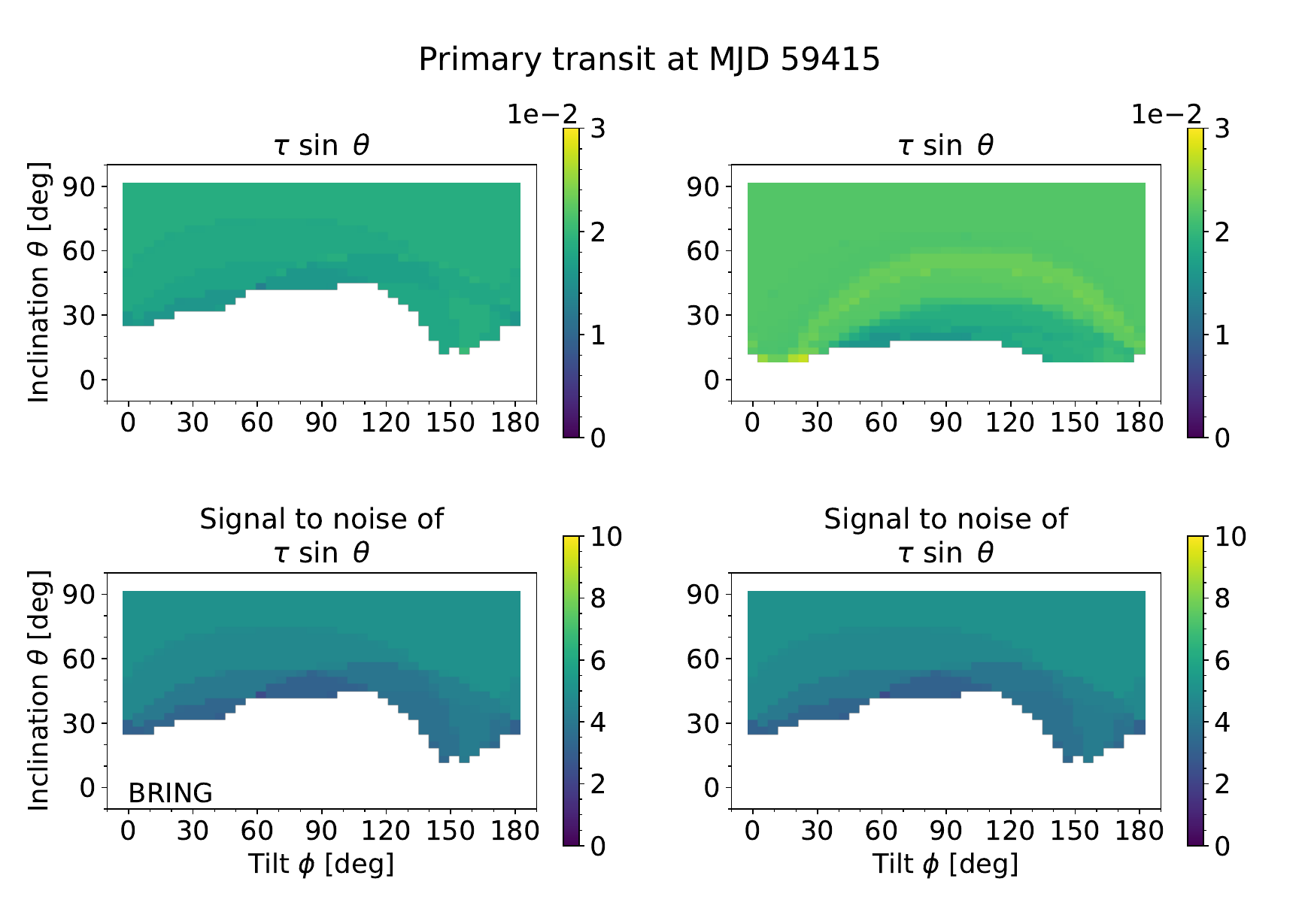}
  \caption{The optical depth (upper subpanel) and SNR (lower subpanel) for a CPD fit at varying tilts and inclinations for the Hill sphere transit at MJD 59415 and a circumplanetary radius of 0.3 $r_{\text{Hill}}$ (upper panel) and 0.6 $r_{\text{Hill}}$ (lower panel). 
    White mark the areas with no photometry to constrain the fit and only bRing has temporal coverage (Fig.~\ref{fig:photometric_transits}) during this Hill sphere transit. 
A confident CPD orientation corresponds to a single upper optical depth limit with a high SNR as seen in the upper left plots of Fig.~\ref{fig:model_timestamps}. 
The CPD fit at 0.3 $r_{\text{Hill}}$ has no upper optical depth and the fit at 0.6 $r_{\text{Hill}}$ shows a loosely constrained disk orientation.} 
    \label{fig:59415diskfit_3datasets_taumass}
\end{figure*} 

The second primary transit at MJD 59415 shown in Fig.~\ref{fig:59415diskfit_3datasets_taumass} has been solely covered by bRing.
The midpoint of the transit is better constrained than at MJD 58195 with an uncertainty of 8 days.
A similar pattern within the optical depth values at 0.6 $r_{\text{Hill}}$ are seen between both transits, where the same conclusion applies. 
Despite bRing showing the highest SNR and largest $\tau$ values at MJD 59415, no constraints are placed on any plausible CPD.
The differing results among the instruments at different epochs and CPD radii do not yield compelling evidence to constrain the inclination, tilt, or radius of a possible CPD.

\begin{table*}[h!] 
    \caption{The dominant grain size $D_{\text{min}}$ and upper mass limit $M_{\text{CPD}}$ for the primary transits at different CPD radii.}
    \begin{subtable}{.5\linewidth}
      \centering
\begin{tabular}{ccc}
  \hline
  \multicolumn{3}{c}{Transit at MJD 58195 } \\
  \hline
  \hline
& 0.3 $r_{\text{Hill}}$ & 0.6 $r_{\text{Hill}}$\\
\hline
 \hline
$D_{\text{min}}$  [$\mu m$]  & 18.1    &25.6\\
$M_{\text{CPD}}$ [g]&   $10^{19.1}$  & $10^{19.6}$ \\
 \hline
\end{tabular}
    \end{subtable}%
    \begin{subtable}{.5\linewidth}
      \centering
      
        \begin{tabular}{ccc}
 \hline
 \multicolumn{3}{c}{Transit at MJD 59415} \\
 \hline
 \hline
& 0.3 $r_{\text{Hill}}$ & 0.6 $r_{\text{Hill}}$\\
 \hline
$D_{\text{min}}$ [$\mu m$]  & 18.1    & 25.6\\
$M_{\text{CPD}}$ [g]&   $10^{20.9}$  & $10^{21.7}$ \\
 \hline
\end{tabular}
    \end{subtable} 
    \label{tab:uppermass}
\end{table*} 

Fig.~\ref{fig:taumass_58195} and \ref{fig:taumass_59415} in the Appendix show the upper mass limit for MJD 58195 and 59415, respectively, using Eqn. \ref{eq:mass_limit}.
Table \ref{tab:uppermass} shows the mean dust size and upper mass limit constrained by the upper limits of $\tau$.
The grain size was computed to be 18.1 and 25.6 $\mu$m, for 0.3 and 0.6 $r_{\text{Hill}}$ respectively for both transits. 
The upper mass limits at MJD 58195 were determined to be $10^{19.1} $ g and $10^{19.6}$ g for 0.3 and 0.6 
$r_{\text{Hill}}$ respectively and at MJD 59415 $10^{20.9}$ g and $10^{21.7}$ g for 0.3 and 0.6 $r_{\text{Hill}}$ respectively.
We place an upper mass limit of the order $10^{22}$ g in the Hill sphere of \picc{}.

Possible ways of improving future studies of \pic{} include (i) constraining the midpoint by improved orbital fits of \picc{}, and (ii) running a more general fitting code with the midpoint $t_b$ as a free parameter \citep[e.g. \texttt{emcee}; ][]{Foreman-Mackey_2013}.
The grain size $D_{\text{min}}$ used to calculate the upper limit of the CPD mass assumes a homogeneous distribution and constant grain size in the disk. 
It is possible that there is no disk and the material has accreted onto the planet or exomoons, giving that the system is at the end of the planet formation process, or the disk could have a low obliquity and does not transit the star. 

\section{Conclusions}\label{sect:conc}
Following the Hill sphere transit of \picb{}, we have carried out an analysis on the data used in \citet{Kenworthy_2021} to search for any material transiting the star in the Hill sphere transits of \picc{}.

BRITE, bRing and BRITE partially covered \picc{} during the two primary transits at MJD 58195 and 59415 and a secondary transit at MJD 58707.
All transits and additional discontinuous coverages were fitted to a model of the 1981-like event to search for compelling evidence of a similar detection.
None were found except the detection at MJD 58690 could be a possible 1981-event caused by an unknown event.
Fluctuations with a magnitude increase to $m \sim$ 0.01 over a span 5-15 day by ASTEP and bRing were detected, but a lack of similar detections in the BRITE data suggest terrestrial sources of photometric noise that were not removed by the photometric pipelines.

Potential CPD tilts and inclination were probed for radii of 0.3 and 0.6 $r_{\text{Hill}}$ using a model from \citet{Kenworthy_2021}.
The model's sensitivity to the midpoint of the offset was analysed, showing partial detections of a CPD with the correct orientation for up to three days offset.
Taking the sensitivity of the model into account, the optical depth at the primary transits MJD 58915 and MJD 59415 were computed.
The former transit has full coverage from bRing and BRITE, however only the fit using the BRITE data yielded feasible positive optical depth.
The latter transit was only covered by bRing.
The results are inconsistent for the different instruments at MJD 58195 and inconsistent among both transits.

The poorly constrained midpoints of the transits and the highly sensitive model yield inconclusive results. 
With an orbital period of 3.0 years, additional transits of the Hill sphere can be monitored and circumplanetary material searched for.
Most notably, the PLATO satellite \citep{2025ExA....59...26R}, due for launch at the end of 2026, will monitor \pic{} as part of its LOPS2 field for at least two years. 
The next transit of the \picc{} Hill sphere will therefore be seen in two broadband colours with the PLATO array with space satellite millimagnitude precision, enabling the detection of material at much lower optical depths around this planet.

\section{Data availability}

We are committed to open science and have made the data, reduction scripts, and plots in this paper available on an open-source basis\footnote{Available at \url{https://github.com/mkenworthy/BetaPiccHS}}.
All code is provided under a Berkeley Software Distribution (BSD) 2-Clause `Simplified' license.
\begin{acknowledgements}
We thank all who contributed to the bRing project, acquiring data from BRITE, bRing and ASTEP.
We thank Sam Mellon for his contributions to the bRing project during the course of his PhD.
Part of this research was carried out in part at the Jet Propulsion Laboratory, California Institute of Technology, under a contract with the National Aeronautics and Space Administration (80NM0018D0004).
\end{acknowledgements}

\bibliographystyle{aa_url}
\bibliography{bib}

\onecolumn
\begin{appendix}
\label{sec:appendix}
\section{The 1981 event}
\label{sec:appendix_1981_event}
The continuation of Fig.~\ref{fig:cross_correlation} is shown in Fig. ~\ref{fig:cross_correlation2}.
\begin{figure}[ht]
\script{06_1981_fit.py}
    \centering
        \includegraphics[trim={0cm 0cm 0 11cm},clip,width=0.9\textwidth] {figures/1981_fits.pdf}
    \caption{Fig.~\ref{fig:cross_correlation} continued}
  \label{fig:cross_correlation2}
\end{figure}

\section{Upper mass limits}

\begin{figure*}[h!]
    \script{09_CPD_model_mass_58195.py}
    \centering
    \includegraphics[width=0.9\textwidth]{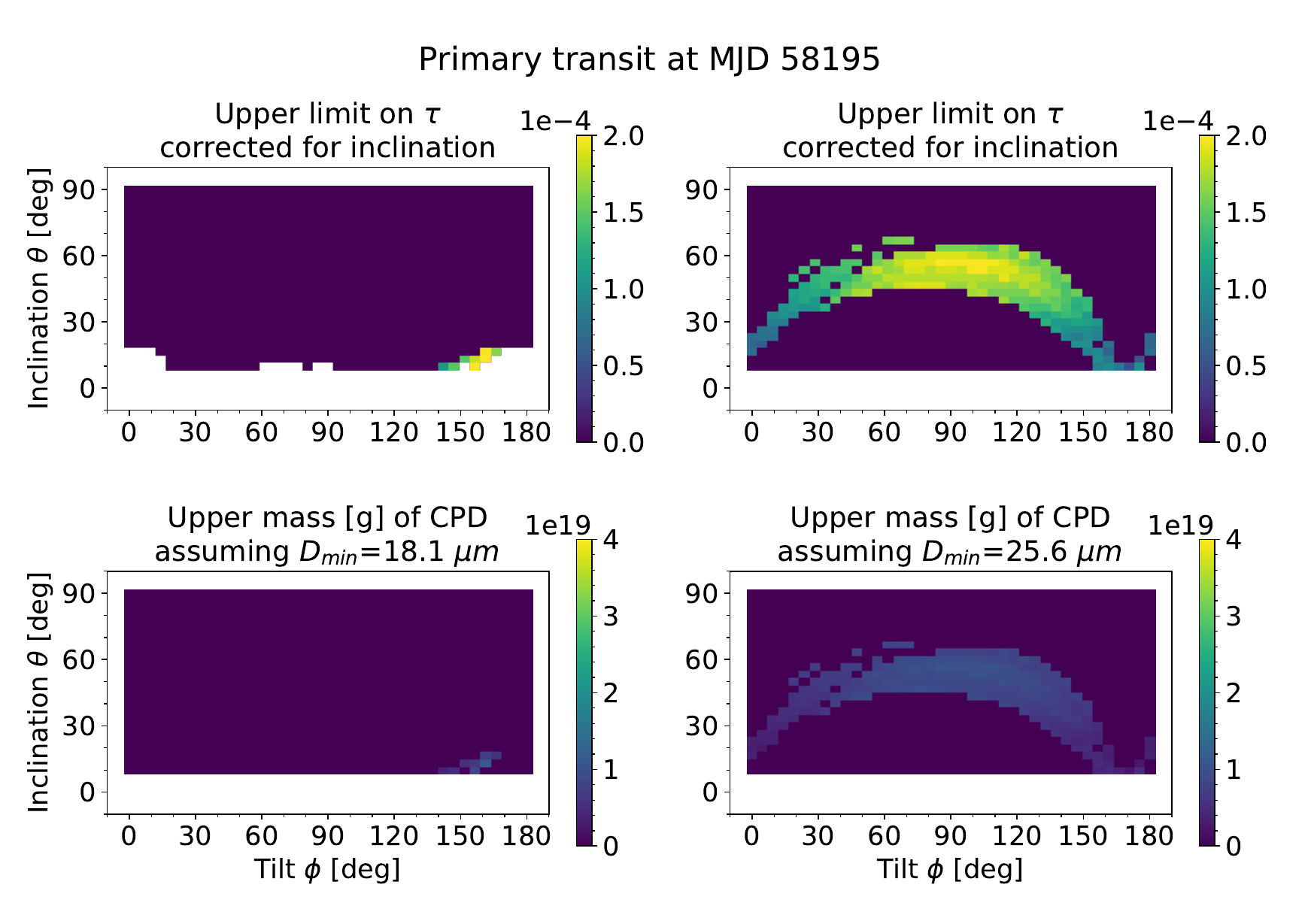}
\caption{The upper optical depth limit corrected for inclination (top panels) and the corresponding upper mass limit (lower panels) for MJD 58195 at 0.3 $r_{\text{Hill}}$ (left panels) and 0.6 $r_{\text{Hill}}$ (right panels).   
A confident CPD disk orientation and upper mass limit result from a single upper optical depth as seen in the upper left plots of Fig.~\ref{fig:model_timestamps}.
The BRITE values from Fig. \ref{fig:58195diskfit_3datasets_taumass} were used with no compelling disk orientation and their computed grain sizes are 18.1 $\mu$m and 25.6 $\mu$m, with upper mass limits of $10^{19.1}$ and $10^{19.6}$ g, respectively. }
    \label{fig:taumass_58195}
\end{figure*}

\begin{figure*}[h!]
    \script{09_CPD_model_mass_59415.py}
    \centering
    \includegraphics[width=0.9\textwidth]{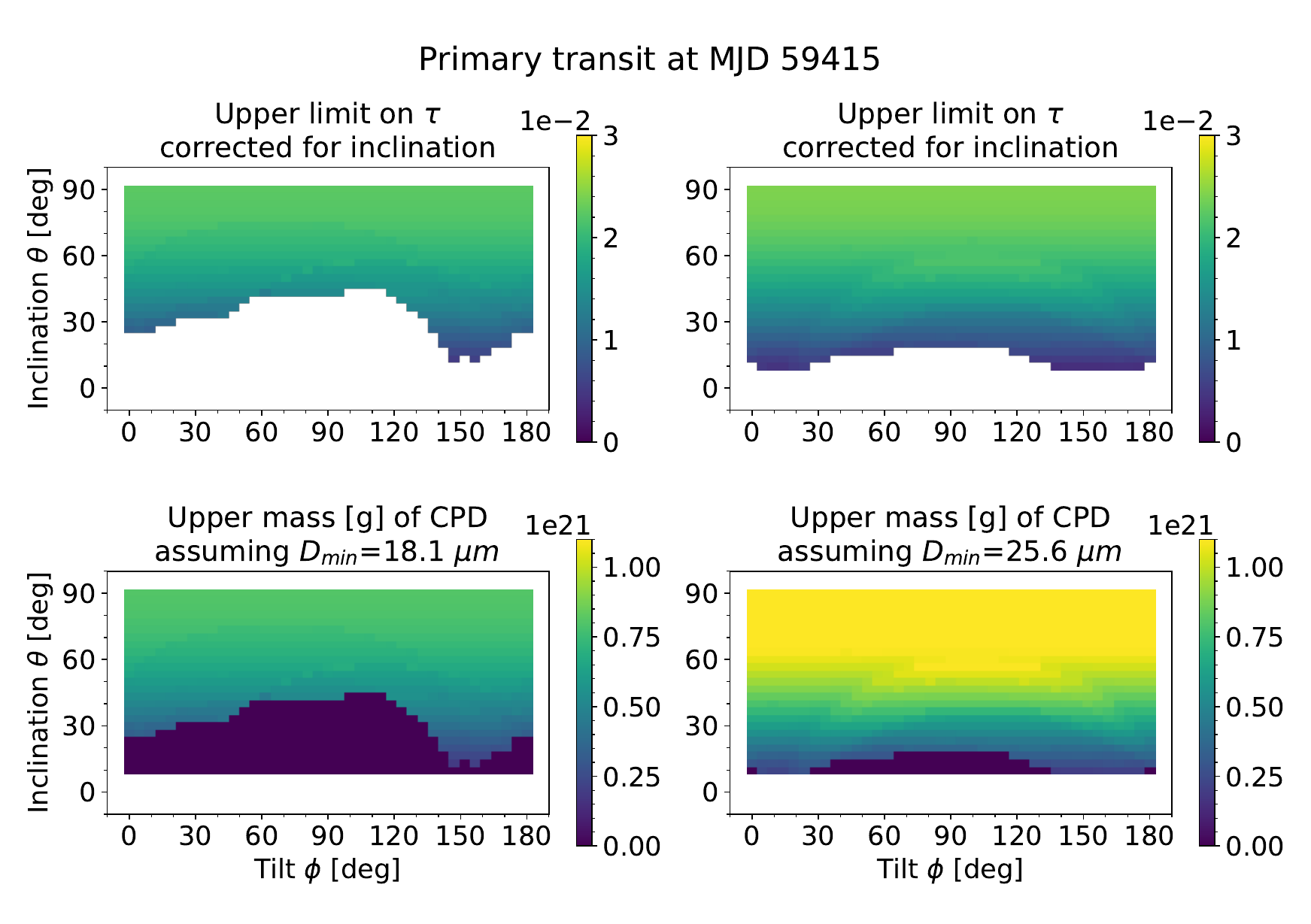}
\caption{
Similar to Fig.~\ref{fig:taumass_58195}.
The bRing values from Fig. \ref{fig:59415diskfit_3datasets_taumass} were used with no compelling disk orientation and their computed grain sizes are 18.1 $\mu$m and 25.6 $\mu$m, with upper mass limits of $10^{20.9}$ and $10^{21.7}$ g, respectively. }
\label{fig:taumass_59415}
\end{figure*}

The upper mass limits shown in Fig.~\ref{fig:taumass_58195} and \ref{fig:taumass_59415} are computed using Eqn. \ref{eq:mass_limit} and the optical depths of Fig.~\ref{fig:58195diskfit_3datasets_taumass} and Fig.~\ref{fig:59415diskfit_3datasets_taumass}.
Fig.~\ref{fig:taumass_58195} shows the upper limit of $\tau$ corrected for inclination and the upper mass limit at 0.3 $r_{\text{Hill}}$ and 0.6 $r_{\text{Hill}}$ for MJD 58195 using only BRITE data. 
The upper mass limit is found when a single upper value of $\tau$ within the grid exists, however there is none.
At 0.3 $r_{\text{Hill}}$ no constraints can be placed, whereas at 0.6 $r_{\text{Hill}}$ the disk orientation and upper mass limit can be loosely constrained.
Fig.~\ref{fig:taumass_59415} shows the upper mass limit at MJD 59415 from bRing allows all possible orientations of the disk, thus the two different transits do not agree with each other.
Despite the many uncertainties contributing to the results, the upper mass limits of the grain size are listed in Table \ref{tab:uppermass}.

\end{appendix}
\end{document}